\documentclass[twocolumn]{webofc}

\usepackage[varg]{txfonts}   
\usepackage{graphicx}
\graphicspath{{./images/}}

\begin{document}
\title{Teaching young learners a foreign language  via tangible and graphical user interfaces}

\author{\firstname{Heracles} \lastname{Michailidis}\inst{1} \and
        \firstname{Eleni} \lastname{Michailidi}\inst{1}   \and
        \firstname{Stavroula} \lastname{Tavoultzidou}\inst{1} \and
        \firstname{George F.} \lastname{Fragulis}\inst{1}
        \fnsep\thanks{\email{gfragulis@uowm.gr}}
}

\institute{Laboratory of Robotics, Embedded and Integrated Systems, \\Dept. of Electrical and Computer Engineering,\\
	University of Western Macedonia, Hellas}

\abstract{%
  The use of tangible interfaces in teaching has been proved more effective, user -friendly and helpful in collaborative learning departments, when compared to traditional teaching approaches. In particular, the tangible interface "Makey Makey"is  a modern tool that enhances collaboration between pupils, with positive results in education, despite the limited research done on this interface so far. "Makey Makey" succeeds in motivating and engaging young learners in the learning process, showing better performance and scoring results. In addition, its use in teaching has been shown to benefit the learning process in every age learning group.The development and use of such an innovative teaching/learning approach  helps young learners perceive the educational process in a different way and assimilate new cognitive fields more effectively. Moreover, educators profit as well, as they can eliminate difficulties and  teach more efficiently using examples based on their teaching approach, while enhancing young learners’ parallel skills  as well. This study will confirm previous research results stating that assimilation of new concepts is easier with tangible interfaces than with graphical ones, as well as that young learners participating in the survey have shown significant progress in knowledge acquisition when compared to their prior knowledge.
}
\maketitle

\section{Introduction}
\label{intro}
The evolution of educational process and the need for assimilation of the teaching material render learner’s active involvement an essential component in the teaching process. Such an involvement may entail discussions between the learner, the educator and or other learners,   as well as exchange of views and ideas concerning the teaching/learning process. Learning is a key factor in  knowledge acquisition, therefore, participatory teaching, where all members of a class have an active role and present their ideas and knowledge through activities related to the lesson, is a teaching approach much more effective than traditional ones. In this study we use the Makey Makey, a tangible technology device, which, with the provision of direct and interactive environments, can innovate the teaching process offering learners the opportunity to get in touch with programming and design \cite{Lee2014} . Created by Jay Silver and Eric Rosenbaum in 2010, Makey Makey  is an electronic board which can help us  convert everyday objects into computer input devices.  

\subsection{The 5E Instructional Model}
The 5E Instructional Model, introduced by  \cite{bybee1990science}, can be used in science education for syllabus design and implementation. It comprises five cognitive stages of learning, i.e. engagement, exploration, explanation, elaboration and evaluation. 
\begin{figure}[h]
	\includegraphics[width=0.5\textwidth]{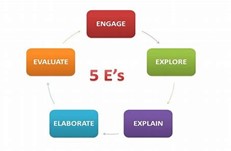}
	\caption{The 5E Instructional Model}
	\label{5E}       
\end{figure}

According to \cite{bybee1997achieving}-(p.176)   “using this approach, students redefine,reorganize, elaborate, and change their initial concepts through self-reflection and interaction with their peers and their environment. Learners interpret objects and phenomena, and internalize those interpretations in terms of their current conceptual understanding". The 5E teaching model can be used, either for integrated learning programs, for educational modules, or individual lessons. The five phases of learning of the 5E Instructional Model are discussed in detail as follows:\\
\textbf{Engagement}. This is a student-centered phase, during which the learner is motivated to learn  more about the upcoming topic.The teacher, on the other hand, focuses on assessing learners’ prior knowledge, as well as identifying possible misunderstandings. During this phase students may brainstorm, ask or formulate questions about the topic with an aim, on behalf of the teacher, to create interest or generate curiosity. The use of short activities closely relating to students’prior knowledge and the new syllabus will facilitate the acquisition of new concepts and enhance the learning process.\\
\textbf{Exploration} is also a student- centered phase during which the learner through active exploration experiences learning. The teacher acts as a facilitator or consultant encouraging students to work cooperatively with other peers in a learning context without his/her direct instruction. Thus, the promotion of process skills, such as observation, questioning, investigation, testing predictions, hypothetisizing, communication is encouraged. The importance of this phase lies on the fact that students have a “hands-on” experience before formal instruction begins.\\
\textbf{Explanation} is a phase more teacher-centered, as the teacher is called to serve as a facilitator who will ask the students to describe and discuss their experience during the exploration phase. Moreover, students can pose questions, ask for clarifications of misconceptions emerged during the engagement or exploration phase, express their explanations and ideas about a concept, before the teacher attempts to explain it. During this phase the teacher can provide formal definitions, notes, labels, as well as may integrate computer software programmes, video, visual aides to facilitate students’ understanding. Thus, students will be able describe the main concepts both to the teacher and their peers.\\
\textbf{Elaboration}. During this phase students are encouraged to apply  new acquired knowledge, as well as acquiring new skills, with an aim to develop deeper and broader understanding of the concepts. Thus, they are called to check for understanding with their peers, exchange ideas and information, design experiments or integrate their knowledge and skills to other disciplines.  Integration of technology in activities elaboration, i.e., web based research, WebQuests is therefore encouraged in this phase.\\
\textbf{Evaluation} is the phase of assessment, but differs from assessment in conventional science lessons because both formal and informal  assessment approaches can be employed to evaluate students’ learning. Viewing assessment as an ongoing process, teachers can observe whether their students can apply new concepts and skills, or have modified their way of thinking. Evaluation may also include self-assessment or peer-assessment, a quiz, an exam or a writing assignment \cite{Cardak2008}. 
In conclusion, the 5E Instructional model can result in students’ better performance and scoring, as it facilitates the process of learning  new concepts and alters the way students view the teaching process. In addition, it can facilitate educators provide a student-centered learning environment, as it  is more flexible for syllabus design and lesson planning.

\section{Using Computer Applications in Education}
Information and Communication Technology (ICT) or Information Technology(IT) has totally changed the way people think work and live, introducing the “knowledge society” we are called to live in. New curricula  have to be developed therefore, to prepare young learners for the new era, in which technology plays a dynamic role. The integration of  (ICT) /(IT) in conventional teaching can facilitate not only the delivery of lessons, but also the learning process. The use of  IT/ICT can alter the learning context from teacher-centered to student -centered, as students are familiar with technology, therefore more motivated, and can be more actively involved in the learning process. 
Incorporating visual stimuli in the teaching process can help young learners understand theoretical concepts, as they can make connotations between their prior knowledge of the world and the new concepts taught. Young learners’ prior knowledge is an essential factor in the learning process, therefore the teacher should take it into account during syllabus design and teaching planning in order to achieve the desired results. Games and intense stimuli, when used as pedagogical tools, can enhance students’ interests and actively insert them into the teaching and learning process. Moreover, the encouragement of socialization and synergies between peers  will be of crucial importance, as well. 
Regarding the learning outcomes, young learners will take initiatives, develop cognitive competence, new skills, creative and complex thinking, will assimilate new knowledge more effectively and perform and score better \cite{Shaffer2005}.

\subsection{Tangible user interfaces}
Tangible user interfaces (TUIs) are interfaces that allow a user interact with digital graphics and information through the physical world. Specifically, by TUIs, physical and tangible objects are manipulated with the use of technologies in an environment to control the functions of the computer \cite{Antle2007}. A computer system can become an extension of ourselves if we use an object and control it with it. Thus, the object acquires substance in the execution of the control process and loses its real structure, as it becomes a channel of communication between the user and  the digital world. Real and digital can be interlinked, either by combining an everyday object and a robotic hand, or by using a physical hand and a graphic, or by using a physical hand and a real object. Essentially in the interface, the user is asked to use his body in order to interact with the computer and thus understand the concepts taught. TUIs are made up of features that make them important, i.e., the accuracy of the input devices, the access to the interaction, the use of the hands, the tangible interaction and the connection between the control of the physical object and the use of its digital representation. It is important to point out that the use of tangible interfaces does not require computer skills, as during interaction, physical objects and the user's body are only used \cite{Horn2007}. This reduces the difficulty of learning a computer environment and the user can acquire the desired knowledge through the easy use of the tangible user interface . The digital media provided should be interactive, interesting, student-friendly, easy to use and suitable to the age group it is addressed to, in order to help children develop their creativity, learn new things, enhance collaboration with their peers.

\section{What is "Makey Makey"}
The "Makey Makey" is a USB device consisting of alligator cables and clips through which it is connected to conductive materials that function  as  keys that control and motion graphics in the computer  \cite{2012}. The board sends closed-loop electrical signals to the computer either via the mouse or the keyboard and does not require special software, programming or operation skillset \cite{Collective2012}.

\begin{figure}[h]
	\includegraphics[width=0.5\textwidth]{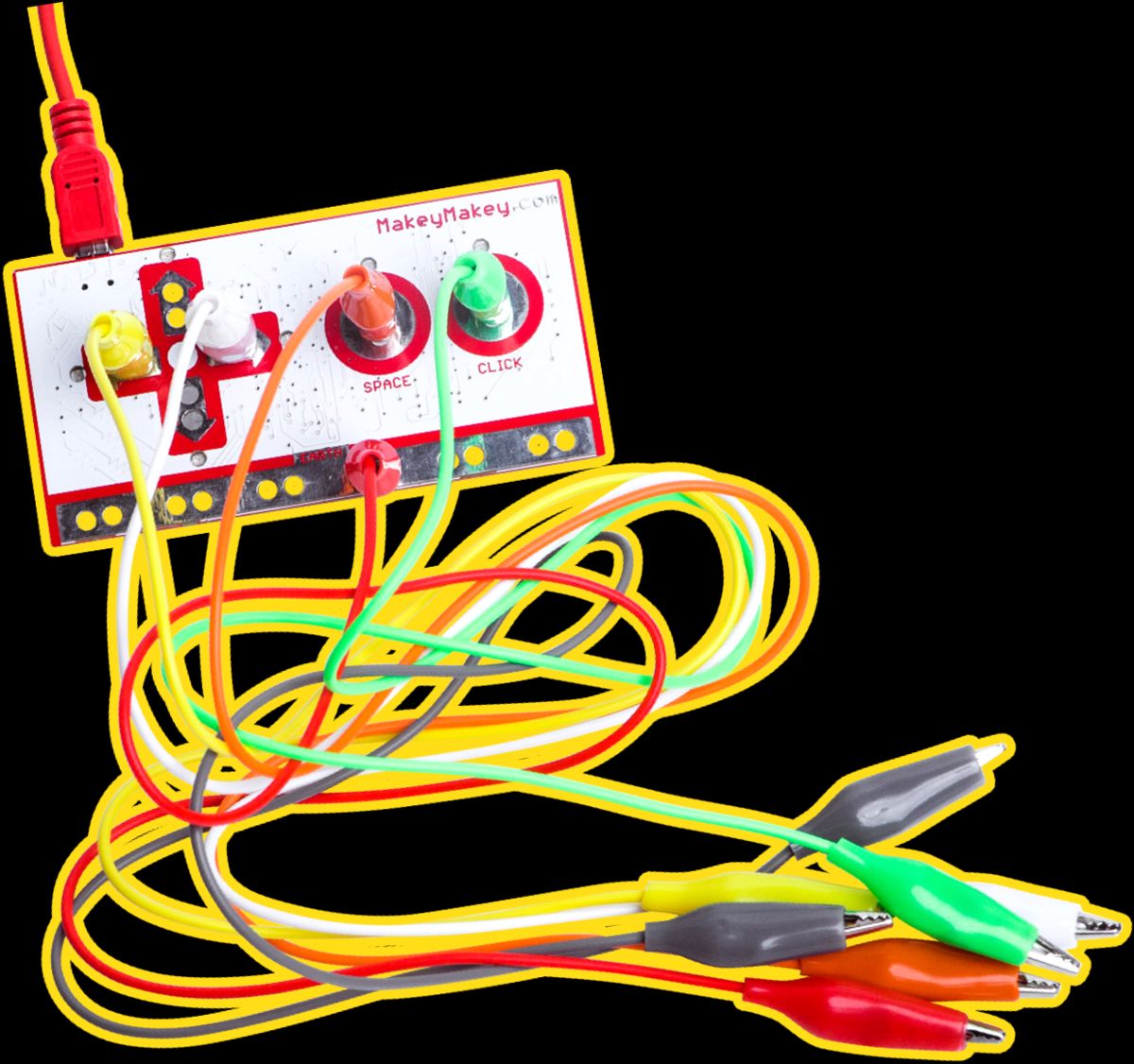}
	\caption{Makey-Makey}
	\label{makey}       
\end{figure}

\begin{figure}[h]
	\includegraphics[width=0.5\textwidth]{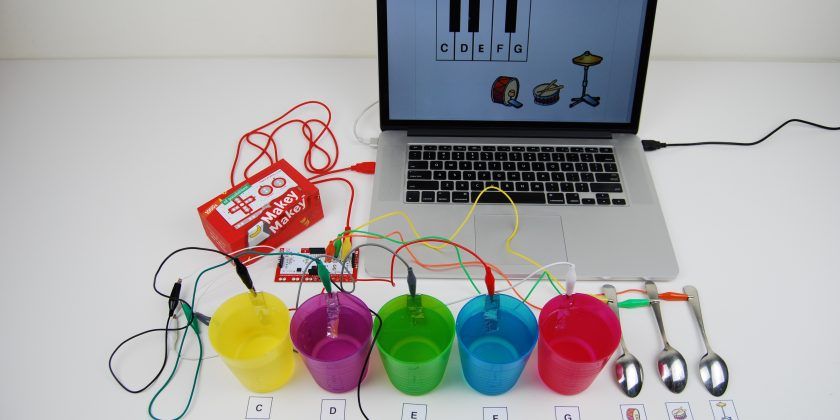}
	\caption{Makey-Makey connected to a PC}
	\label{makey2}       
\end{figure}

"Makey Makey" comes in many forms depending on the connection tools and the devices where it is connected to. It consists of:
\begin{itemize}
	\item 'MaKey Makey' board, which has six inputs on the front corresponding to four arrow keys, a space bar, a mouse input
	\item the USB cable that connects the board to the computer
	\item four alligator clips, which connect the board with a key-object
\end{itemize}
"Makey Makey" works as a keyboard/mouse that transmits commands to the computer. The alligator clips are easy to use, suitable for immediate repositioning tests by users, and are placed in holes in the circuit board. After "Makey Makey" is connected to the computer, the user can make circuits \cite{Kafai2015}. To close an electrical circuit, the user must hold or touch a ground wire connected to the "Makey Makey" as it connects another wire to an object, provided it is conductive \cite{GarciaRuiz2016}. In this way, the circuit closes and the computer receives it as a keystroke \cite{Resnick2013}. With the objects connected to the board, the active program of the computer is controlled and the tangible interface is created. Creating interfaces is an easy process that simply requires exploration and testing with electrically conductive objects .All objects that conduct electricity even on a low scale can be used. Some categories that can be used are foods, dried plants, plasticine, smooth pencil drawings, metals, and the human body \cite{Makey}.
"Makey Makey" simply helps a computer program to work by touching on conductive objects. The main goals of the creation of "Makey Makey" are \cite{Collective2012}:

\begin{itemize}
	\item Ease of use for beginners
	\item Computer interaction with physical objects
	\item Compatibility with all software
	\item No planning required
	\item It is easy to assemble and connect to the computer
	\item It is attractive due to its ease of use
	\item It takes minimal time to learn how to operate and connect to a computer and materials
\end{itemize}
 
\subsection{Advantages}
\label{sec-2}
Tangible interfaces provide an interesting environment that allows for interaction and can improve the school learning environment. Tangible interfaces widely help simplify teaching by providing tools to better understand inaccurate lesson terms through interactive images \cite{Antle2013}. Also, they encourage physical participation using the body, help children to explore and discover new skills, and to express views and questions about their participation in the educational process \cite{Cramer2015}. Combining tangible interfaces with digital imaging or haptic applications \cite{Kokkonis} is a new process that can help students seek out more knowledge and reconsider existing ones more effectively \cite{Rogers2009}. TUIs are also used for collaboration in creating games and educational environments where children through cooperation with their classmates are encouraged to take initiative and develop parallel skills. \cite{Schneider2010},\cite{Schreiber2013}. Tangible interfaces can be used by students with learning disabilities, young age groups and beginners \cite{Zuckerman2005}. It is open-source hardware under CC license and can be utilized with other open-source software applications. In recent years our research team has developed a number of applications using open-source programming languages and tools such as PHP, MySQL and WordPress \cite{Fragulis2018}, \cite{Lazaridis2016}, \cite{Lazaridis2019}, \cite{Michailidi2020}, \cite{Papatsimouli2020}, \cite{Skordas2017}.

\subsection{Disadvantages}
Tangible interfaces, despite the features they give to the learning process, have difficulties in their use, which are mainly related to the lack of information about the use of objects and the steps that must be taken to achieve a goal. Another difficulty is that they are designed for an audience with a different background of knowledge.

\section{Using the tangible interface Makey- Makey in classroom}
According to \cite{Morrison2007}, "The quality of a research is not only enhanced or weakened by the correct methodology and the correct use of the research tools, but also by the appropriateness of the sample selected". In surveys where the sample consists of many people and it is not possible to measure one hundred percent of the population due to time, costs, etc., a certain number of people is used as a representative sample for the research.
In the present research, the sample consists of small pupils of the Katranitsa Foreign Language Center of Drama, Greece (aged 3 to 6 years). The sample is 60 pupils who were divided into 2 groups, group 1 which was taught animal names using MIT Scratch software on a computer, and group 2 which was also taught animal names using the interface developed based on " Makey Makey ".
For both groups of the pupils, a "match the card" game was developed with the help of the MIT Scratch software, in which the student is asked to match the cards with pictures of animals that he has in the respective sounds in a frame. Each pupil, by pressing specific areas of the box, hears the name of the animal in English and the sound it makes and is asked to find and place the matching card. He can listen to the pronunciation as many times as he/she wants. Regarding the way of interface with the application,  group 1 was using a computer and a mouse, and the pupil was asked to drag the cards with pictures of animals in the appropriate place, and then he/she would be able to hear the confirmation sound.

\begin{figure}[h]
	\includegraphics[width=0.5\textwidth]{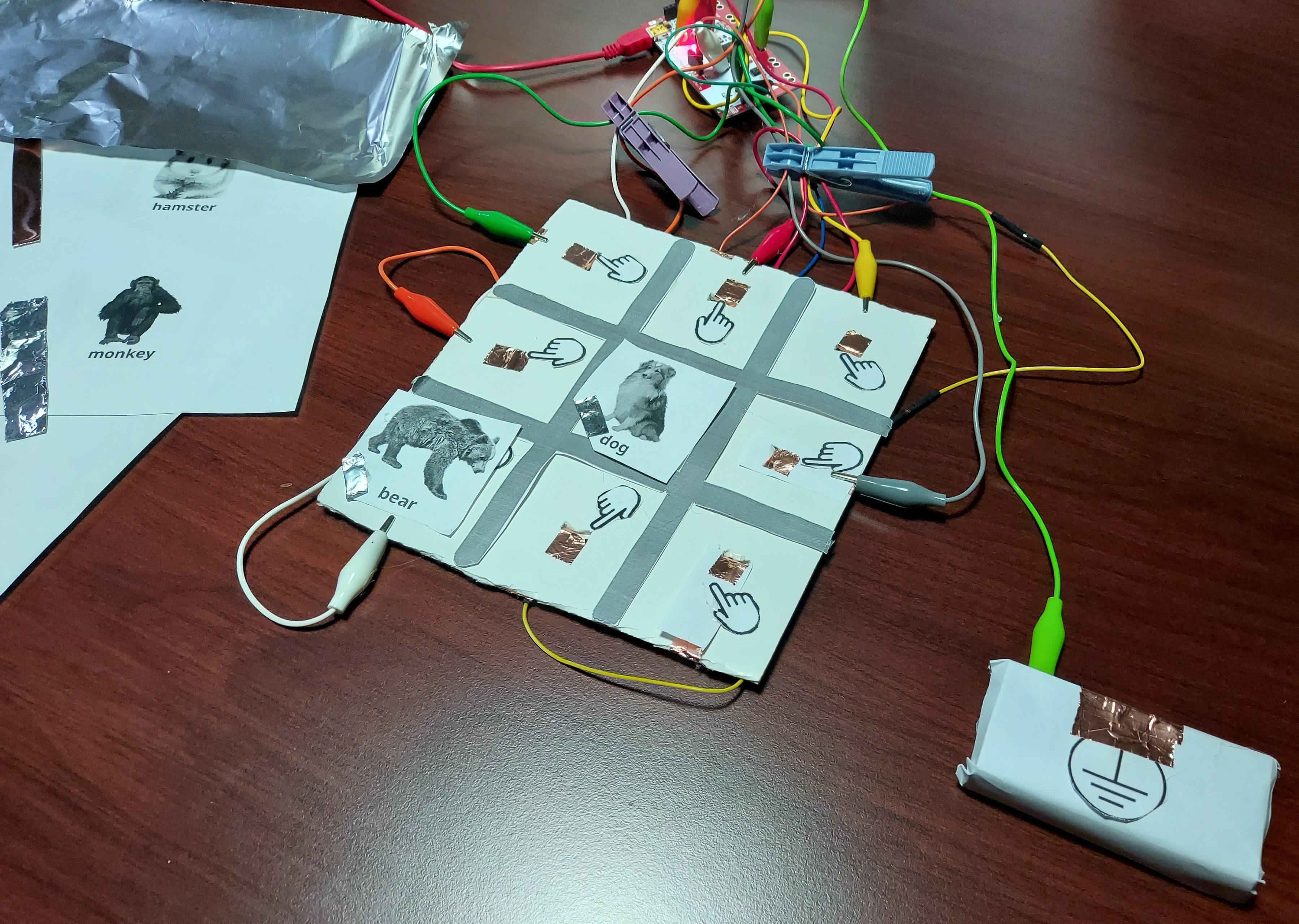}
	\caption{Makey project}
	\label{makey}       
\end{figure}

In group 2, which used the tangible interface, the pupil was given conductive aluminum sheets as well as sheets of paper with pictures of animals which he was asked to cut with scissors and stick the picture to the aluminum creating his conductive card. Next, he was asked to place the cards he made on a grid. By placing a card in one place, he could hear the pronunciation of the word and the sound of the animal.

\subsection{Duration of Research}
The research was conducted simultaneously because it is based on responses of population groups that were investigated during the same period. The duration of the research was six months, from June 2020 to the end of November 2020.  More specifically, the topic of the research was set in the context of the development of innovative teaching in foreign language learning. This process took four months in total. The experimentation in the operation of "Makey Makey" and the development of English language teaching courses for Greek-speaking children lasted the remaining two months. The tests took place for each group of children for one teaching hour (45min), however, it was necessary to make a presentation to get acquainted with "Makey Makey" and "Scratch" which also lasted one teaching hour as well. One week later we've made another test for assessment.

\subsection{Research results}
 
\subsubsection{Word Knowlegde}
\begin{figure}[t]
	\includegraphics[width=0.5\textwidth]{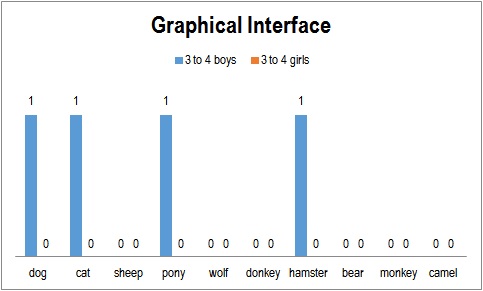}
	\caption{Children age 3 to 4 - Graphical Interface}
	\label{fig-1}       
\end{figure}

\begin{figure}[!h]
	\includegraphics[width=0.5\textwidth]{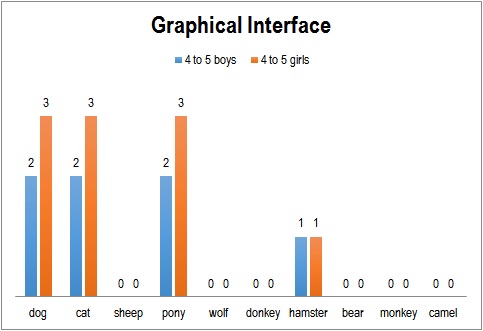}
	\caption{Children age 4 to 5 - Graphical Interface}
	\label{fig-2}       
\end{figure}

\begin{figure}[h]
 	\includegraphics[width=0.5\textwidth]{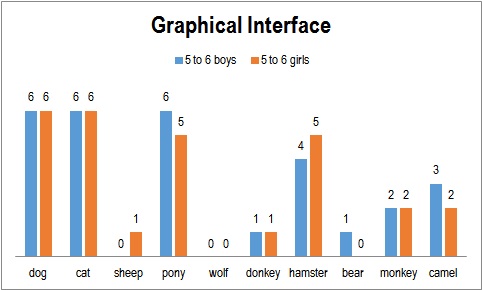}
 	\caption{Children age 5 to 6 - Graphical Interface}
 	\label{fig-3}       
 \end{figure}

Question: Which words do you know in English from the following (cards shown in random order) dog, cat, sheep, pony, wolf, donkey, hamster, bear, monkey, camel
The results are shown in figures (\ref{fig-1})-(\ref{fig-3}):\\

\textit{First group (Graphical Interface):}

\newpage
\textit{Second group (Tangible Interface):}
  
 \begin{figure}[h]
 	\includegraphics[width=0.5\textwidth]{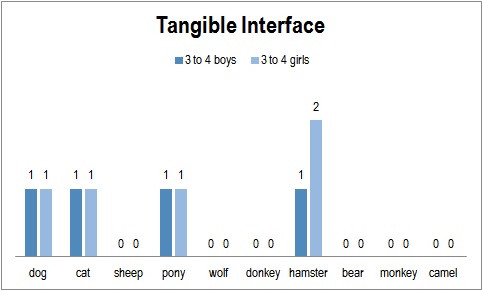}
 	\caption{Children age 3 to 4 - Tangible Interface}
 	\label{fig-4}       
 \end{figure}

\begin{figure}[h]
	\includegraphics[width=0.5\textwidth]{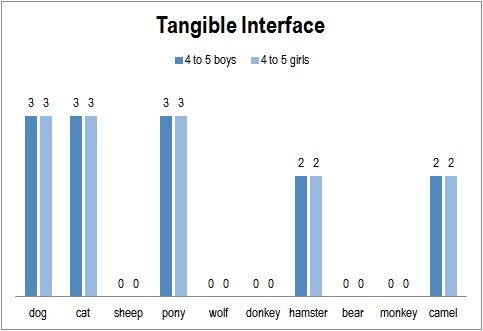}
	\caption{Children age 4 to 5 - Tangible Interface}
	\label{fig-5}       
\end{figure}

\begin{figure}[h]
	\includegraphics[width=0.5\textwidth]{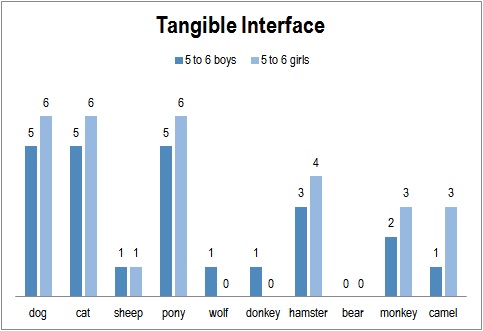}
	\caption{Children age 5 to 6 - Tangible Interface}
	\label{fig-6}       
\end{figure}

\newpage
\subsubsection{Question about using Graphical/Tangible Interfaces before}
Question: Have you done a scratch and/or makey makey lesson in the past?
(Figure (\ref{table2})):

\begin{figure}[h]
	\includegraphics[width=0.5\textwidth]{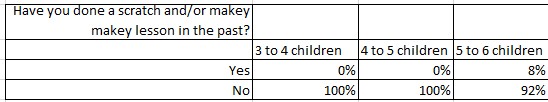}
	\caption{Have you done before a similar lesson}
	\label{table2}       
\end{figure}

\newpage
\subsubsection{Assesment after one week}
We asked the children again after a week period : Which words do you know in English from the following (cards are shown in random order) dog, cat, sheep, pony, wolf, donkey, hamster, bear, monkey, camel?\\

\textit{First group (Graphical Interface):}

\begin{figure}[h]
	\includegraphics[width=0.5\textwidth]{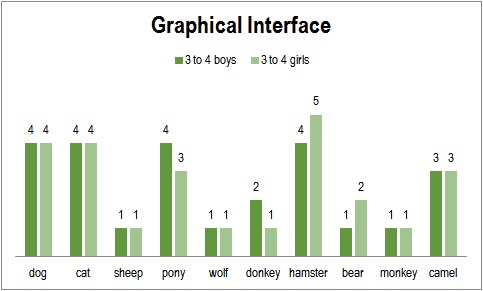}
	\caption{Graphical Interface -1}
	\label{fig-10}       
\end{figure}

\begin{figure}[h]
	\includegraphics[width=0.5\textwidth]{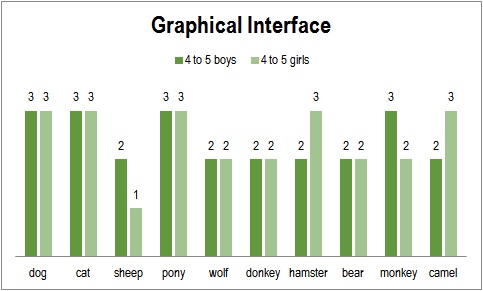}
	\caption{Graphical Interface -2}
	\label{fig-11}       
\end{figure}

\begin{figure}[h]
	\includegraphics[width=0.5\textwidth]{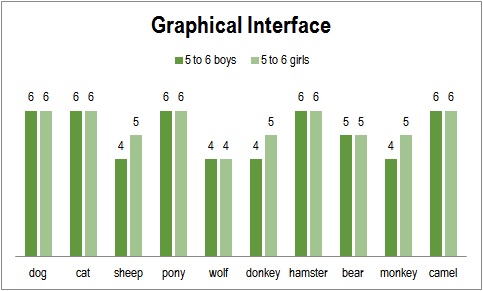}
	\caption{Graphical Interface -3}
	\label{fig-12}       
\end{figure}

\newpage
\textit{Second group (Tangible Interface):}

\begin{figure}[h]
	\includegraphics[width=0.5\textwidth]{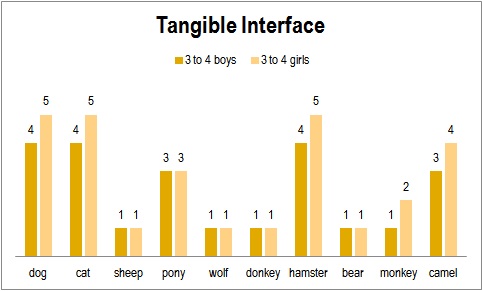}
	\caption{Tangible Interface -1}
	\label{fig-13}       
\end{figure}

\begin{figure}[h]
	\includegraphics[width=0.5\textwidth]{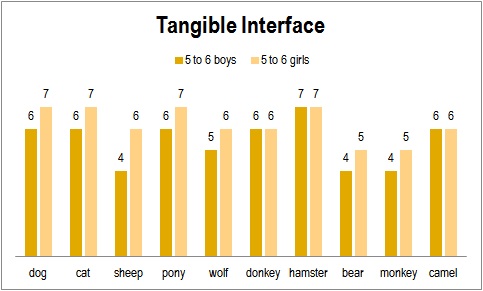}
	\caption{Tangible Interface -2}
	\label{fig-14}       
\end{figure}

\begin{figure}[h]
	\includegraphics[width=0.5\textwidth]{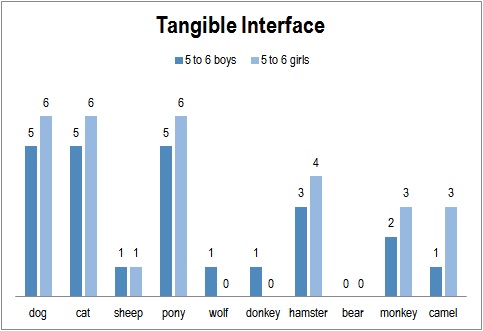}
	\caption{Tangible Interface -3}
	\label{fig-15}       
\end{figure}

\subsubsection{Evaluation}

\begin{figure}[h]
	\includegraphics[width=0.5\textwidth]{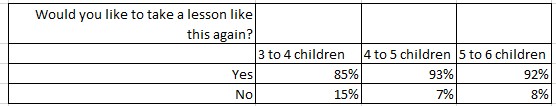}
	\caption{Would you like to do it again}
	\label{table3}       
\end{figure}

In order to evaluate the validation of the interfaces used in this research and how they  affect the students’ learning pace we give them the following questionnaire Figures (\ref{table1})-(\ref{table3}):

\begin{figure}[h]
	\includegraphics[width=0.5\textwidth]{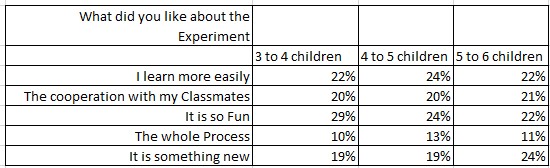}
	\caption{What did you like about the lesson}
	\label{table1}       
\end{figure}

 As a result, the children seemed to be immediately familiar with the new educational way and showed an fast response and improvement of their knowledge through a process that seemed  to be fun. An average score of 90\% stated that they would like to take a similar course again.
 \section{Conclusions}
 According to the above research, it seems that tangible interfaces are more effective for teaching English to similar, age groups of pupils. The pupils who were invited to participate found the process original and fun and stated to a large extent that they would like to repeat their teaching in the same way. The results of the overall research show that the assimilation of new terms is easier through the tangible interfaces than with the graphical interface as well as that the students showed significant progress in the whole process compared to the knowledge they had before the start of the experiment.  
 \subsubsection*{Future Research}
 The limits set in the research and the implementation of the study concerned the extension to more pupils and more time for evaluation . In particular, the translation of the material into other languages and the application in bigger classes were confined. This is something that can be investigated in the future, as well as some other issues related to this study. In detail, the proposals for further research are the following:
 
 \begin{enumerate}
 	\item Expansion of the platform in terms of teaching other languages besides English i.e. French, Italian, Japanese, Chinese or Arabic.
 	\item Expansion of the teaching material.
 	\item Application of the teaching material in a real class learning context with more pupils.
 \end{enumerate}




%

%

%
%

\end{document}